\documentclass{PoS}

\title{Complex Langevin Dynamics in 1+1d QCD at Non-Zero Densities}

\ShortTitle{Complex Langevin Dynamics in 1+1d QCD at non-zero densities}

\author{\speaker{Sebastian Schmalzbauer }
		\thanks{supported by the Frankfurter F\"orderverein f\"ur Physikalische Grundlagenforschung}\hspace{1ex}$^{\ddagger}$\\
		Institut f\"ur Theoretische Physik, Johann Wolfgang Goethe-Universit\"at\\
		Max-von-Laue-Stra\ss e 1 - 60438 Frankfurt am Main - Germany\\
		E-mail: \email{schmalzbauer@th.physik.uni-frankfurt.de}}

\author{Jacques Bloch
		\thanks{supported by the Deutsche Forschungsgemeinschaft via SFB/TRR-55.}\\	
		Institut f\"ur Theoretische Physik, Universit\"at Regensburg\\
		Universit\"atsstra\ss e 31 - 93053 Regensburg - Germany\\
		E-mail: \email{jacques.bloch@ur.de}}

\abstract{We present our results obtained from gauge cooled complex Langevin simulations in 1+1d QCD at non-zero densities in the strong coupling regime with unrooted staggered fermions. For small quark masses there are regions of the chemical potential where this method fails to reproduce correct results. In these parameter ranges we studied the effect of different gauge cooling schemes on the distributions of the fermion determinant as well as of observables.}

\FullConference{34th annual International Symposium on Lattice Field Theory\\
         24-30 July 2016\\
         University of Southampton, UK}

\usepackage{bbm}
\usepackage{braket}
\usepackage{enumitem}

\begin{document}
\section{Introduction}
One of the most demanding problems of computational physics is the sign problem, arising from the fluctuating phase of a complex action. A promising candidate to overcome it could be the complex Langevin method, a stochastic quantization approach \cite{DAMGAARD1987227} first discussed in \cite{PARISI1983393}. However, convergence to correct expectation values is only guaranteed under certain conditions \cite{Aarts:2009uq,Nagata:2016vkn} and therefore one has to use benchmarks for validation with computational costs usually growing exponentially with system size.

Therefore, we study QCD in lower dimensions, in particular 1+1d, where the sign problem is already severe enough for a meaningful test of the complex Langevin method and running benchmarks is still possible with either the subset method \cite{Bloch:2015iha} or reweighted phase-quenched simulations, making it an ideal test model before applying this method to full QCD.

\section{Partition function and Dirac operator}
We consider the strong coupling partition function of a single quark
\begin{equation}
\label{eq:Z}
%	Z = \int \prod_{n,\nu} \mathrm{d}U_{n\nu} \,\, \det D\left(\{U_{n\nu}\}\right)
	Z = \int \mathcal{D}[U] \det D[U],
\end{equation}
using the notation $[U] = \{U_{n\nu}\}$ for simplicity. The two-dimensional staggered Dirac operator
\begin{equation}
\label{eq:D}
	D_{k,l\vphantom{\hat{0}}} = m \delta_{k,l\vphantom{\hat{0}}} + \frac{1}{2} \left[e^{\mu} U_{k0\vphantom{\hat{0}}} \delta_{k+\hat{0},l} - e^{-\mu} U_{l0}^{-1} \delta_{k-\hat{0},l}\right] + \frac{1}{2}(-1)^{k_0}\left[U_{k1\vphantom{\hat{0}}}\delta_{k+\hat{1},l} - U_{l1}^{-1}\delta_{k-\hat{1},l}\right]
\end{equation}
depends on the quark mass $m$, the chemical potential $\mu$ and the gauge link configuration $[U]$. We considered antiperiodic boundary conditions in the temporal direction $\nu=0$, where a unit step in direction $\nu$ is denoted by $\hat{\nu}$.

Presented results are obtained from small lattices $4\times 4$ with lattice spacing $a=1$.

\section{Complex Langevin dynamics}
With Gell-Mann representation of the gauge links
\begin{equation}
\label{eq:GM}
	U_{n\nu} = \exp \left[i \sum_{a=1}^{8} \omega_{n\nu a} \lambda_a \right],
\end{equation}
the discretized CL update step is given by the rotation
\begin{equation}
\label{eq:CL_update}
	U_{n\nu}' = R_{n\nu}U_{n\nu} \qquad\qquad R_{n\nu} = \exp\left[i\sum_{a=1}^{8}\left(\epsilon K_{n\nu a} + \sqrt{\epsilon} \eta_{n\nu a}\right) \lambda_a \right],
\end{equation}
with time discretization $\epsilon$, real Gaussian distributed noise $\eta_{n\nu a}$ of variance 2 and deterministic drift term $K_{n\nu a}$ depending on the chosen discretization scheme, e.g. for action $S = -\log \det D$:
\begin{equation}
K_{n\nu a} = -\frac{\partial}{\partial \alpha}\left.S\left(\left\{e^{i\delta_{n,n'} \delta_{\nu,\nu'}\alpha  \lambda_a} U_{n'\nu'}\right\} \right) \right|_{\alpha=0}
\label{eq:euler}
\end{equation}
for Euler discretization. We used a Runge-Kutta scheme for faster convergence in $\epsilon$. Independent of the discretization, the drift becomes complex for $\mu \neq 0$ and automatically drives the $\mathrm{SU(3)}$ gauge link starting configuration into $\mathrm{SL}(3,\mathbbm{C})$.

There are proofs regarding the correctness of this method under certain conditions, i.e. that the expectation value of an observable $\mathcal{O}$ is the same in the original and complexified theory
\begin{equation}
\label{eq:equivalence}
\braket{\mathcal{O}}= \!\! \int\displaylimits _{\mathrm{SU(3)}} \!\!\! \mathcal{D}[U] \, \det D[U] \, \mathcal{O}[U] = \!\!\!\!\!\! \int\displaylimits _{\mathrm{SL}(3,\mathbbm{C})} \!\!\!\!\!\!\! \mathcal{D}[U] \, P_{\mathrm{CL}}[U] \, \mathcal{O}[U],
\end{equation}
where $P_{\mathrm{CL}} \in \mathbbm{R}$ denotes the sampling probability of the complex Langevin method. See \cite{Aarts:2009uq, Nishimura:2015pba} for problems with
\begin{enumerate}[label=(\Alph*)]
\item large excursions in the imaginary direction\label{excursion}
\item insufficient suppression of singular drifts and singular observables\label{existence}
\end{enumerate}
during the complex Langevin evolution, spoiling the correctness of Eq.~(\ref{eq:equivalence}).

\section{Gauge cooling}
It has been shown first for heavy quarks \cite{Seiler:2012wz} that applying deterministic gauge transformations
\begin{equation}
\label{eq:GT}
	U_{n\nu} \to U_{n\nu}^{(G)}= G_nU_{n\nu}G^{-1}_{n+\hat{\nu}}
\end{equation}
with $G_n \in \mathrm{SL}(3,\mathbbm{C})$ after each Langevin step can cure convergence to wrong expectation values caused by the problems mentioned above. One could think that a gauge invariant theory should not be influenced by (\ref{eq:GT}), however the drift term $K_{n\nu} = \sum_{a} K_{n\nu a} \lambda_a$ transforms as
\begin{equation}
\label{eq:GT_drift}
	K_{n\nu}[U^{(G)}] = G_n K_{n\nu}[U] G_n^{-1},
\end{equation}
and as a consequence the complex Langevin trajectory is altered. A detailed discussion of the applicability of gauge cooling can be found in \cite{Nagata:2015uga}. However, it is not guaranteed that if (\ref{eq:equivalence}) was violated in the first place, it can be restored using this additional method, in particular if \ref{excursion}, \ref{existence} or both are still present.

The method called gauge cooling, since it is usually implemented by minimizing an $\mathrm{SL}(3,\mathbbm{C})$ gauge variant norm $\mathcal{N}$ by gradient descent \cite{Seiler:2012wz}. Two cooling parameters are introduced: \#cooling iterations and cooling stepsize $\alpha$, which has to be chosen adaptively if the gradient of the norm is too large. In the following, we present the effects of gauge cooling using several different norms.

\section{Results with unitarity cooling}
The norm used most for the cooling procedure so far is the unitarity norm
 \cite{Seiler:2012wz}
\begin{equation}
\mathcal{N}_{\mathrm{u}} = \displaystyle{\sum_{n,\nu}} \mathrm{tr} \left[U^{\dagger}_{n\nu}U_{n\nu} + \left(U^{\dagger}_{n\nu}U_{n\nu}\right)^{-1} - 2\right].
\label{eq:unitarity_norm}
\end{equation}
Minimizing (\ref{eq:unitarity_norm}) results in a gauge-equivalent configuration closer to $\mathrm{SU}(3)$ and counters \ref{excursion} during the CL evolution. Sometimes it is also defined without the second term, but we found no qualitative difference in the simulation results if at least 10 cooling steps with $\alpha = \mathcal{O}(0.1)$ were used.

The effect of this unitarity cooling on observables and their distributions was investigated in \cite{Bloch:2015coa} for 0+1d and 1+1d QCD. The main result was that one can observe a discrepancy with the benchmark if the distribution of the fermion determinant includes the origin, i.e. singular drifts are present (compare \ref{existence}), which was accompanied by skirts in the distributions of observables. This behavior was particularly severe around the phase transition for light quarks, e.g. $m=0.1,\;\mu \in [0.1,0.5]$. For smaller and bigger values of the chemical potential and in the case of heavy quarks ($m>1$) for all $\mu$ values, unitarity cooling was able to push the determinant values away from the origin and correct results were obtained.

\begin{figure}
\centering
\includegraphics[width=0.4\textwidth]{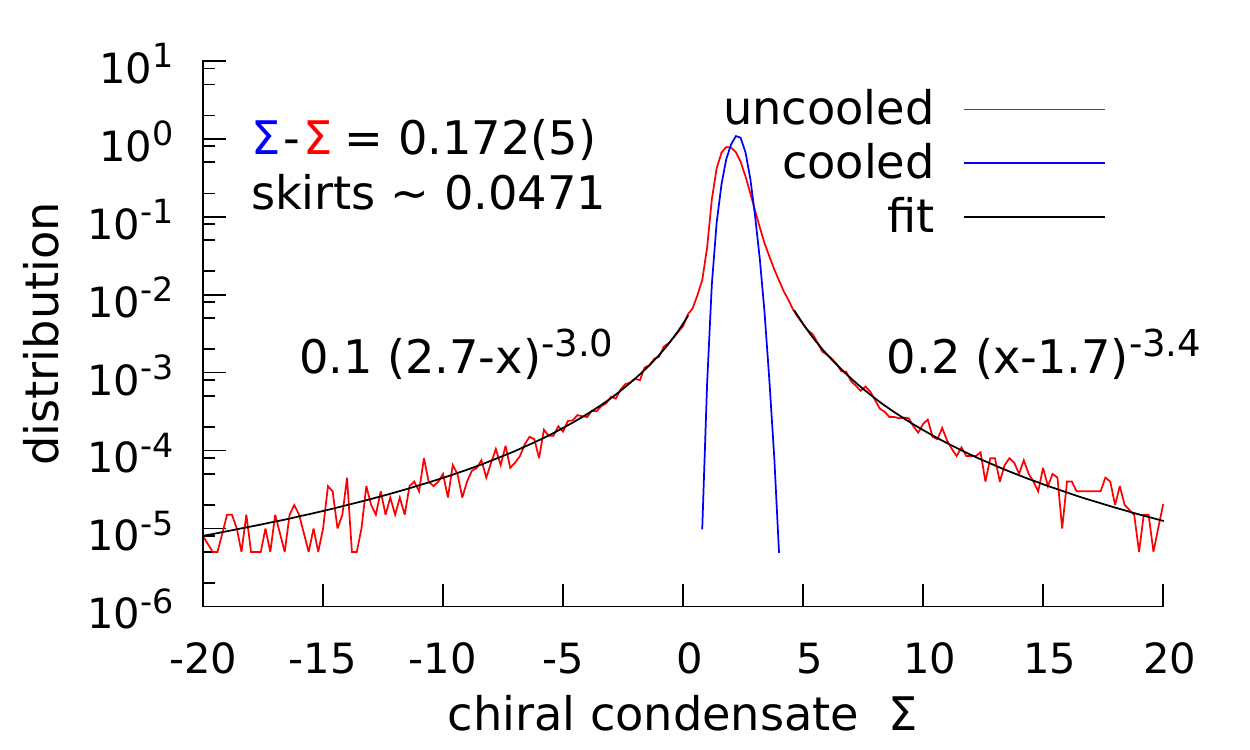}
\caption{Incorrect uncooled simulation with polynomial fit on the skirt fall-off compared to cooled results, which were checked to be correct for the used parameters: $m=0.1$ at $\mu=0.07$.}
\label{fig:skirt_fit}
\end{figure}

A remaining question was the role and origin of the skirts in the observable distribution. To this purpose, we extrapolated their fall-off to clarify whether insufficient cancellations in the tails of the distribution are able to cause the discrepancies. In Fig.~\ref{fig:skirt_fit} this is shown exemplary for the chiral condensate\footnote{We only look at the real part of $\Sigma$ as the imaginary part vanishes on average.}
\begin{equation}
	\Sigma = \frac{1}{V} \; \mathrm{Re} \; \mathrm{tr} D^{-1}
\end{equation}
and we conclude that the skirts themselves are not responsible for the wrong values, but rather the different position of the maximum of the distribution.
Nevertheless, a broad distribution can be used as an indicator for failure of the method, particularly if it drops slower than exponentially.

\begin{figure}
\includegraphics[width=0.32\textwidth]{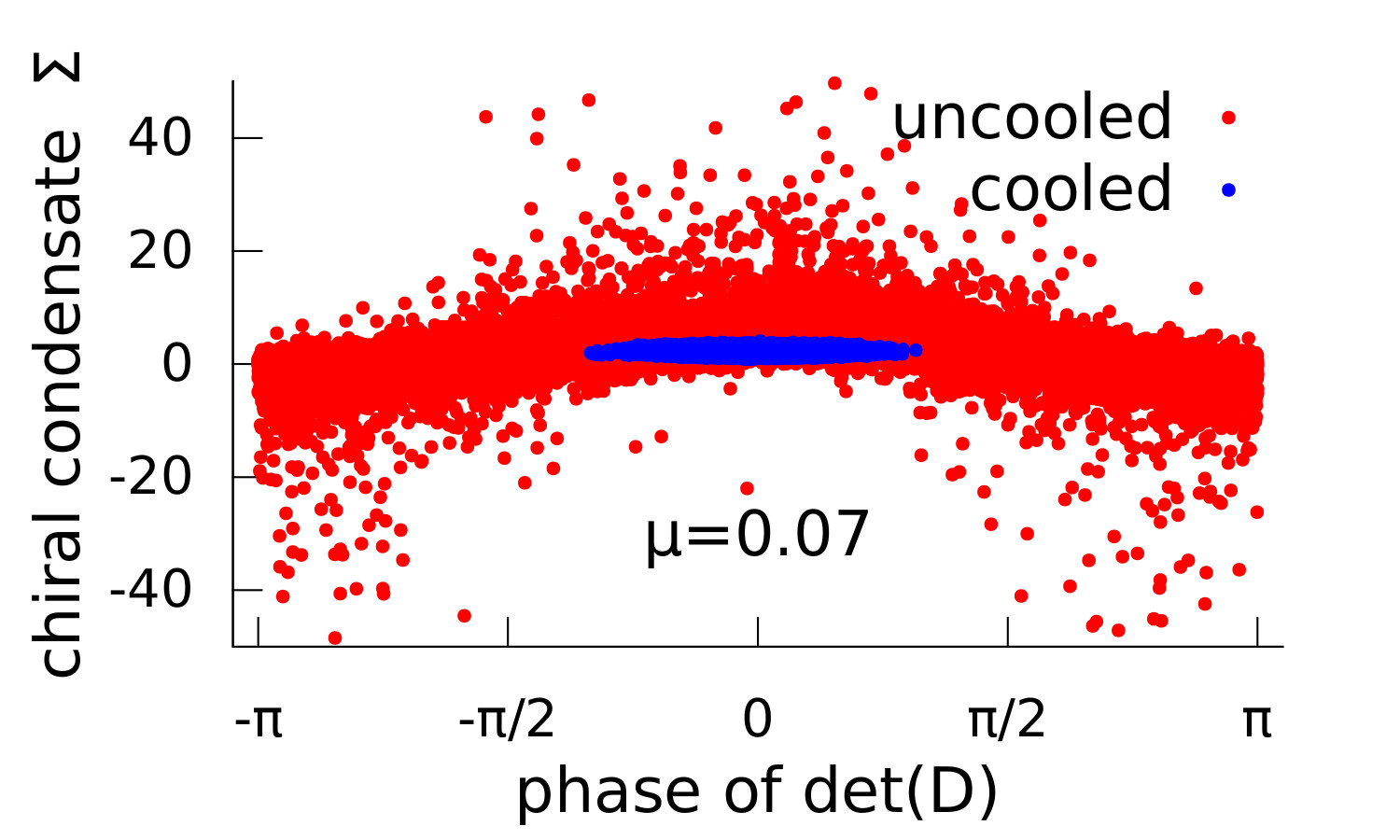}\hfill
\includegraphics[width=0.32\textwidth]{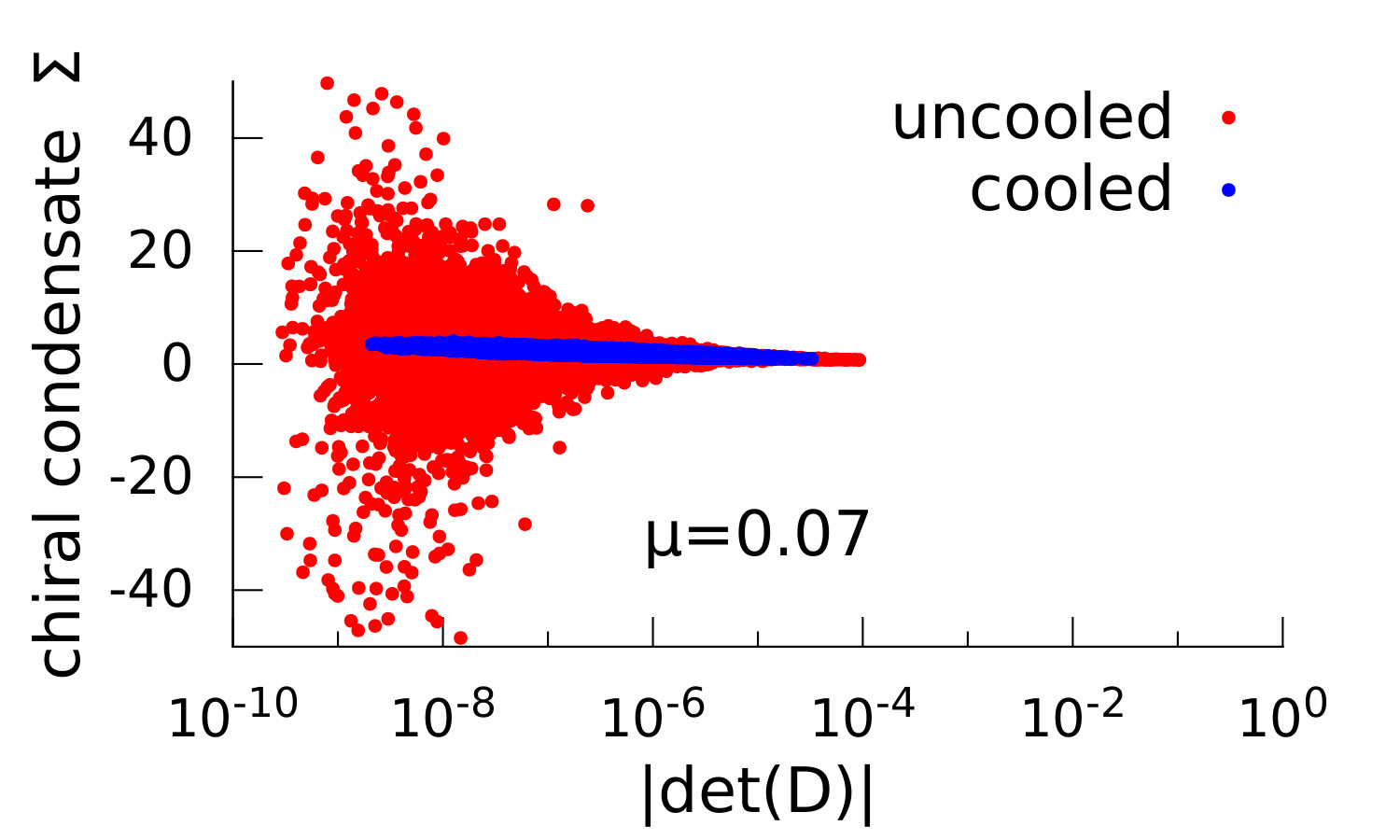}\hfill
\includegraphics[width=0.32\textwidth]{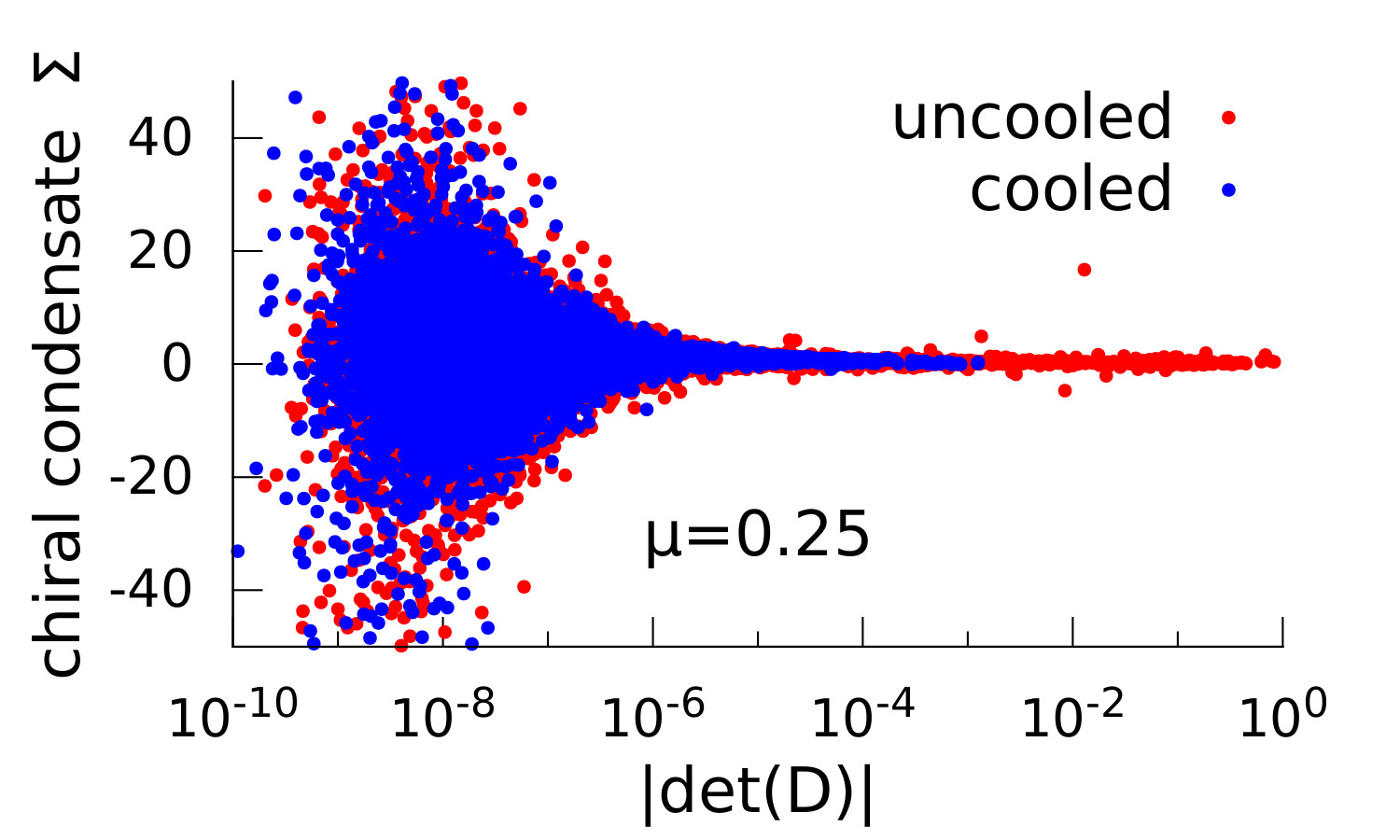}
\caption{Scatterplots to illustrate uncorrelation between skirts an branchcut crossings (left) and different correlation behaviors between skirts and small determinant values (middle, right).}
\label{fig:scatterplots}
\end{figure}

For a better understanding, we wanted to clarify the origin of the skirts and found that they are not related to branch-cut crossings as proposed in \cite{Mollgaard:2013qra}, since they happen independently of the determinant phase, as can be seen in Fig.~\ref{fig:scatterplots}, but correlate with small determinant values corresponding to a singular drift, eg. explicitly for the chiral condensate via
\begin{equation}
\label{eq:correlation}
	\det D = \prod_{i} d_i \quad \Leftrightarrow\quad \Sigma = \frac{1}{V} \sum_{i} d_i^{-1}.
\end{equation}
What is most surprising is that small determinant values can still occur in the absence of skirts, e.g. in cooled simulations at $\mu=0.07$, meaning that the distribution of the eigenvalues $d_i$ of the Dirac operator is also affected by cooling as (\ref{eq:correlation}) still has to hold.

Since cooling the unitarity norm does not perform well enough in some parameter regions, we also studied other cooling schemes, presented in the next chapter.

\section{New cooling schemes}
\subsection{Polyakov cooling}
In 0+1d QCD one can rewrite the theory in terms of a single Polyakov line $P=\prod_t U_{t0}$, also replacing all gauge links in the corresponding unitarity norm:
\begin{equation}
\label{eq:P_norm1d}
	\mathcal{N}_P^{(1\mathrm{d})} = \mathrm{tr} \left [P^{\dagger}P + \left(P^{\dagger}P\right)^{-1} -2 \right].
\end{equation}
This cooling worked extremely well \cite{Bloch:2015coa} and therefore we tried to generalize it for higher dimensions for which we have Polyakov lines $P_x = \prod_t U_{(x,t)0}$ and we define the Polyakov unitarity norm by
\begin{equation}
\label{eq:P_norm}
	\mathcal{N}_P = \sum_{x} \mathrm{tr} \left [P_x^{\dagger}P_x + \left(P_x^{\dagger}P_x\right)^{-1} -2 \right].
\end{equation}
However, since this definition is almost perfectly gauge invariant, minimizing only has an influence on the first timeslice of the temporal gauge links, which was only enough in 0+1d, and therefore the results are not changed perceptively compared to the uncooled case in 1+1d. It could be possible to improve the outcome by modifying Eq.~(\ref{eq:P_norm}) further by inlcuding additional terms in the definition, e.g. Wilson lines $P_t = \prod_x U_{(x,t)0}$ to account for the extra dimension or time shifted Polyakov lines $P_{x,\tau} = \prod_t U_{(x,(t+\tau)\;\mathrm{mod}\; n_t)0}$ as a way to increase the number of affected gauge links by the minimization procedure, but nothing of this has been tested yet.

\subsection{Antihermiticity cooling}
Another idea for cooling is to enhance the antihermiticity of the Dirac operator \cite{Nagata:2015ijn}
\begin{equation}
\label{eq:equation}
D(\mu)^{\dagger} = -D(-\mu),
\end{equation}
which is broken for $\mu \neq 0$ or non-unitary $U_{n\nu}\in \mathrm{SL}(3,\mathbbm{C})$. This property assures that at $\mu=0$ the eigenvalues come in pairs $\pm i \lambda$, resulting in a positive fermion determinant. The corresponding norm is defined as
\begin{equation}
\label{eq:antihermiticity_norm}
\mathcal{N}_{\bar{\dagger}} = \mathrm{tr}\left[\left(D + D^{\dagger}\right)^2\right] \stackrel{}{=} \frac{1}{2}  \displaystyle\sum_{n,\nu} \mathrm{tr} \left[ e^{2\mu \delta_{\nu0}} U_{n\nu}^{\dagger} U_{n\nu} + e^{-2\mu \delta_{\nu0}} \left(U_{n\nu}^{\dagger}U_{n\nu}\right)^{-1} - 2\right] + 12Vm^2,
\end{equation}
where we used Eq.~(\ref{eq:D}) to show that it is very similar to the unitarity norm, only including an additional $\mu$-asymmetry between the first and second term and an irrelevant offset. However, we observed that this is not enough to change the results and determinant or observable distributions compared to unitarity cooling. This is probably due to the fact that unitarity cooling with and without the second term in the norm acts very similarly, as already mentioned, and therefore including a prefactor in front of either of the two terms does not play a significant role at all. 

\subsection{Maximum drift cooling}
The latest proof of correctness of the complex Langevin method \cite{Nagata:2016vkn} only relies on the fact that the probability of drifts with large magnitude is suppressed exponentially. The authors of \cite{Nagata:2016vkn} therefore suggest to reduce the absolute value of the local maximum drift
\begin{equation}
\label{eq:drift_norm}
	\mathcal{N}_d(n) = \max_{\nu}\; \mathrm{tr} \left[K_{n\nu}^{\dagger} K_{n\nu}\right]
\end{equation}
by gauge transformations before the next Langevin update is carried out. For this, we have to use adaptive $\alpha$, because the drift values usually fluctuate quite strongly.

\begin{figure}
\centering
\includegraphics[width=0.32\textwidth]{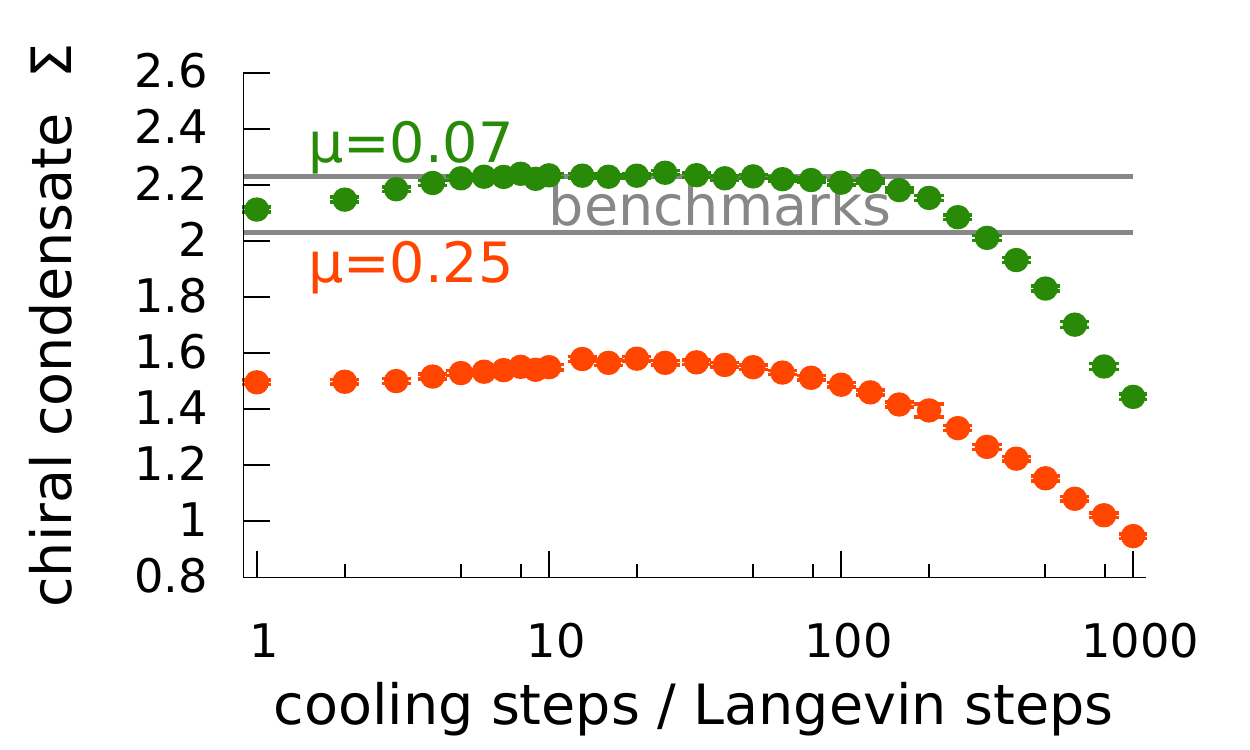} \hfill
\includegraphics[width=0.32\textwidth]{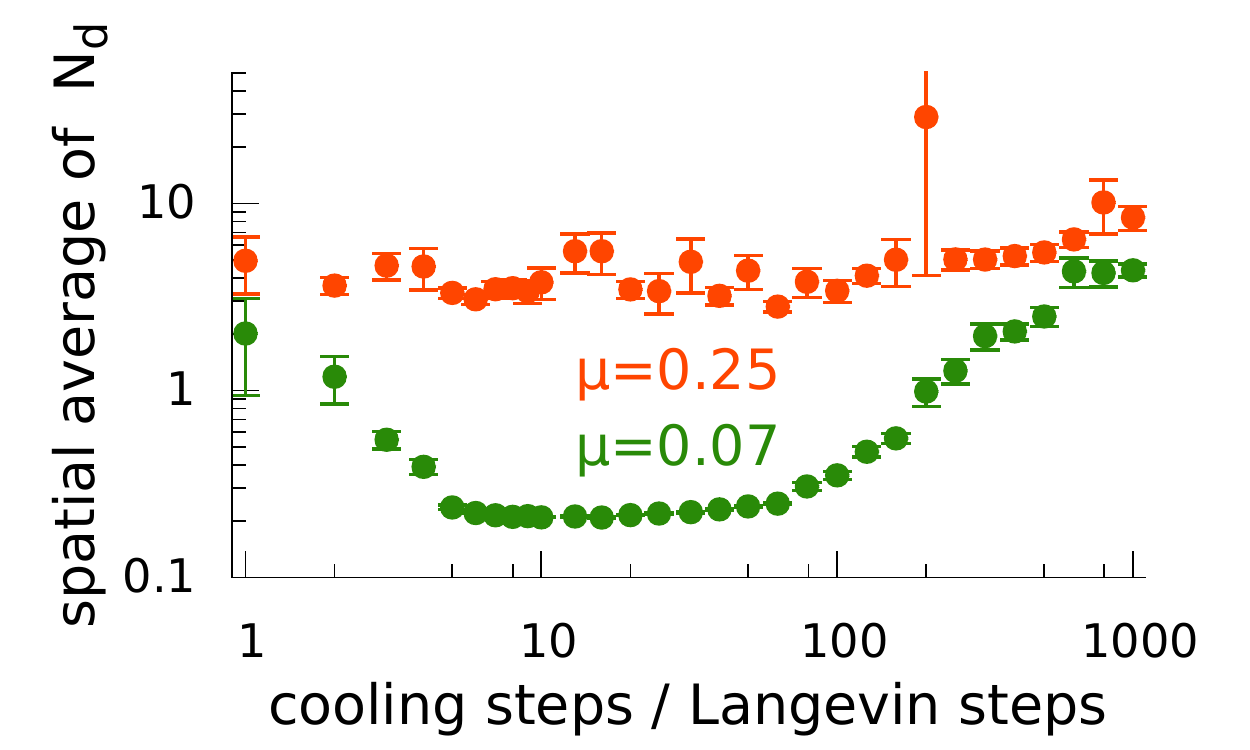} \hfill
\includegraphics[width=0.32\textwidth]{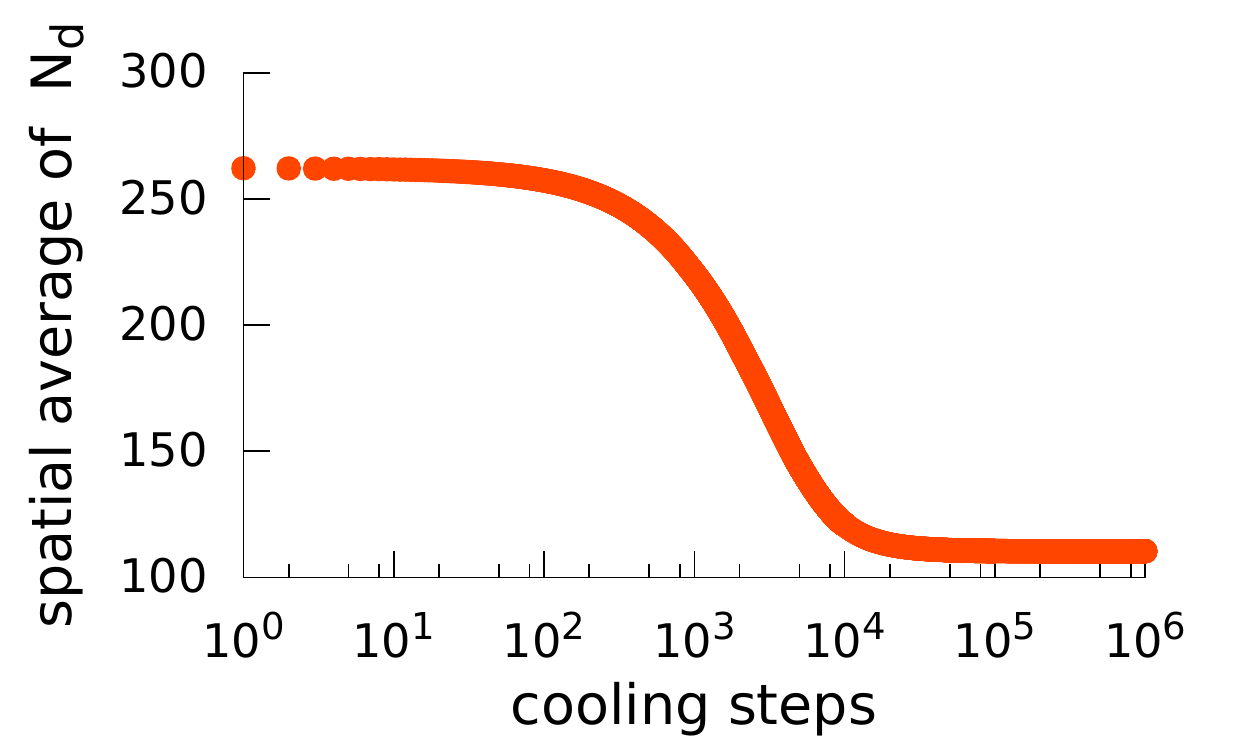}
\caption{Effect of cooling $\mathcal{N}_d(n)$ for small quark mass $m=0.1$ on the chiral condensate (left) and averaged norm (middle), as well as on a single configuration (right).}
\label{fig:drift}
\end{figure}

In contrast to the cooling schemes above, these results somehow depend on \#cooling iterations, although we observe no unusual behavior of the norm (\ref{eq:drift_norm}) when cooling a single configuration excessively, see Fig.~\ref{fig:drift}.
For approximately 10 cooling iterations between the Langevin updates, the method worked best, but not better than unitarity cooling, i.e. we still receive incorrect results for e.g. $\mu=0.25$. Unfortunately, it is also more expensive. We also tried to cool (\ref{eq:unitarity_norm}) and (\ref{eq:drift_norm}) simultaneously, but we could not observe any improvements.

\section{Effect of the gauge action} %ALSO INCLUDE mu=0.07 CASE! FOR THIS, QN SOMEHOW ALSO FAILS
As a supplement to \cite{Bloch:2015coa}, we present some results with gauge action ($S = -\log\det D + S_{G}$)
\[
	S_{G} = \beta \sum_{n}\sum_{\nu < \nu'} \left(1 - \frac{1}{2N_c} \mathrm{tr} \left[U_{n\nu\nu'} + (U_{n\nu\nu'})^{-1}\right]\right), \qquad U_{n\nu\nu'} = U_{n\nu}U_{(n+\hat{\nu})\nu'}(U_{n\nu'}U_{(n+\hat{\nu}')\nu} )^{-1}
\]
for $\beta>0$.
\begin{figure}
\centering
\includegraphics[width=0.4\textwidth]{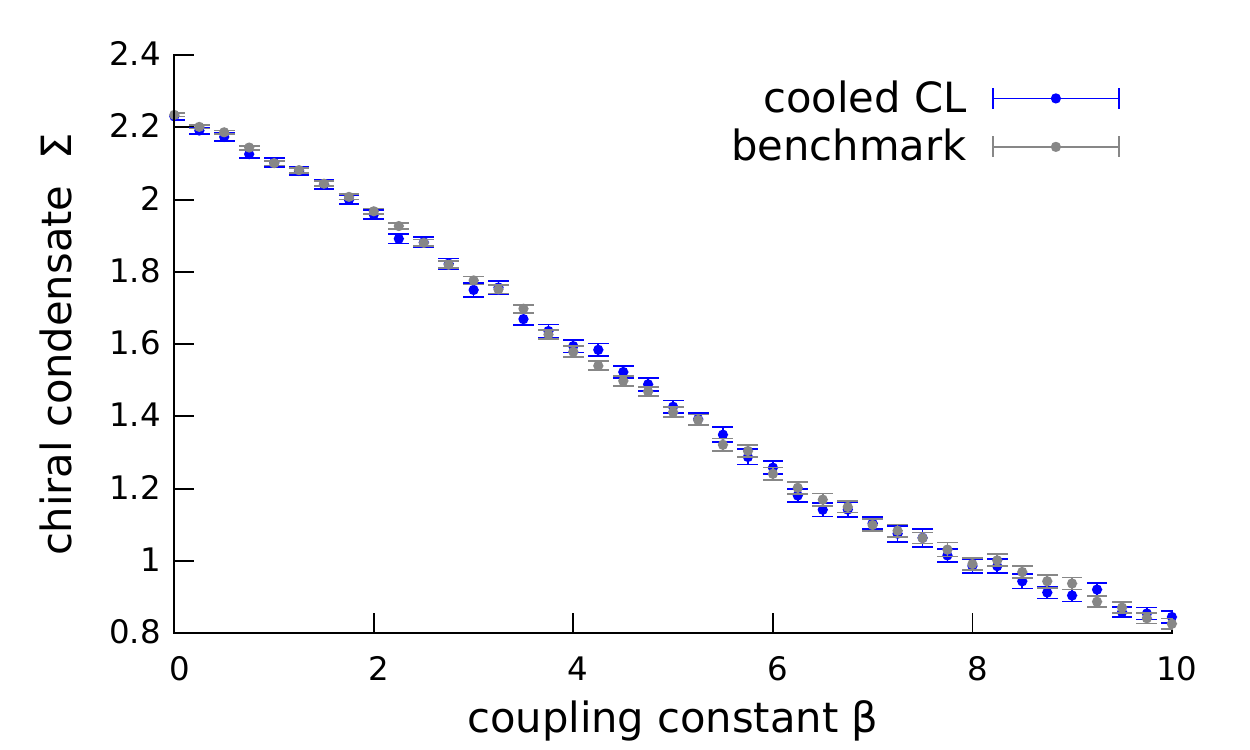}\hspace{1cm}
\includegraphics[width=0.4\textwidth]{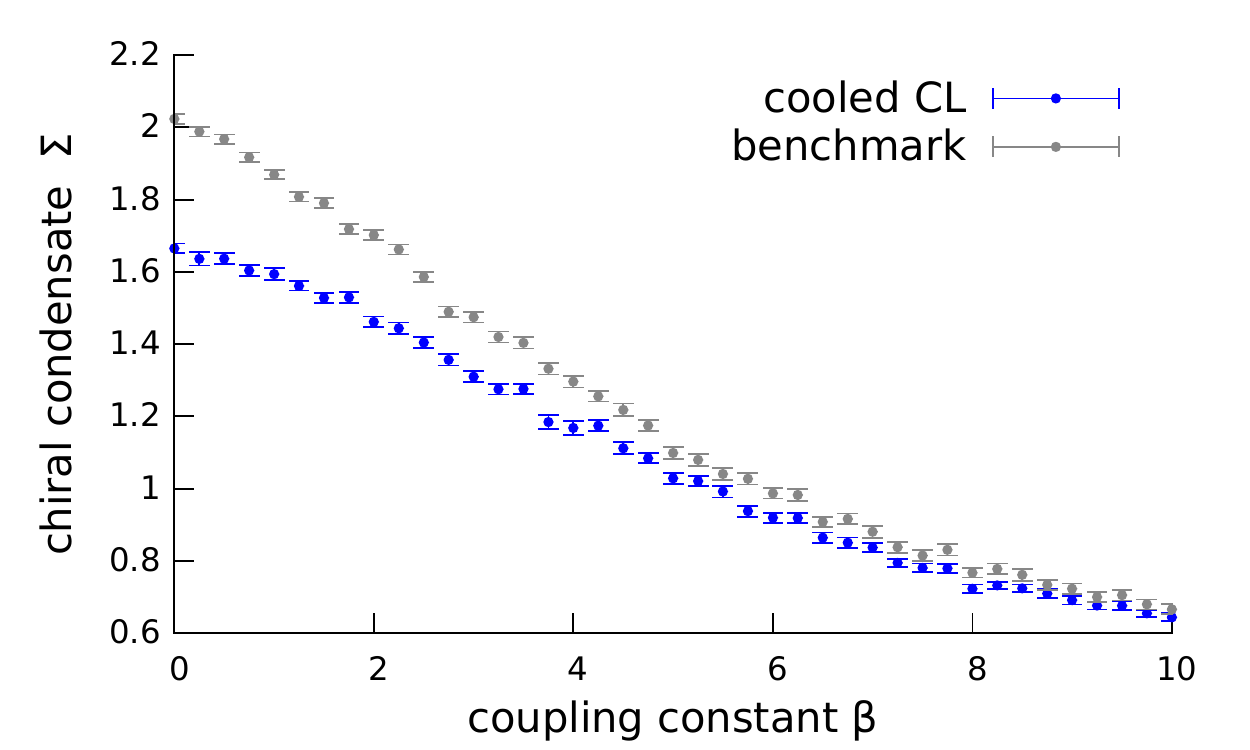}
\caption{Comparison of chiral condensate results from CL and phase-quenched reweighting as a benchmark for $\mu=0.07$ (left) and $\mu=0.25$ (right). In the latter case where CL fails without including the gauge action, increasing $\beta$ has an improving effect and the results slowly converge to the correct ones.}
\label{fig:beta}
\end{figure}
The positive influence on incorrect results by increasing $\beta$ can be seen in Fig.~\ref{fig:beta}. Unfortunately, also the computational effort scales linearly with $\beta$, as the stepsize $\epsilon$ has to be reduced in order to keep the statistical error constant.

\section{Conclusions}
Gauge cooled complex Langevin simulations produce correct results for heavy quarks. However, for light quarks, all considered gauge cooling schemes are insufficient to retrieve the correct values in a significant range of the chemical potential around the phase transition, as the validity conditions of the complex Langevin method are still not met.

Considering the gauge action improves the results for high $\beta$, but the problem at strong coupling still remains.

However, it should be possible to increase the applicability range of the CL method further by combining it with other methods such as reweighting \cite{JBtalk, JBpaper} or Taylor expansions from correct trajectories at $\mu > 0$.
\bibliography{literature}
\bibliographystyle{JHEP}
\end{document}